\documentclass[10 pt]{article}
\usepackage{geometry}
\usepackage{amssymb}
\usepackage{amsmath}
\usepackage{graphicx}
\usepackage{bm,color}
\usepackage{epstopdf}
\begin{document}
\title{Verification of Gyrokinetic codes: theoretical background and applications.}

\author{Natalia Tronko$^{1}$, Alberto Bottino$^{1}$, Tobias G\"orler$^{1}$, Eric Sonnendr\"{u}cker$^{1}$, Daniel Told$^{1}$ 
\\ and Laurent Villard$^{2}$,
\\
$^{1}$Max-Planck-Institut f\"{u}r Plasmaphysik,  85748 Garching, Germany,
\\
$^{2}$ Swiss Plasma Center, Ecole Polytechnique F\'ed\'erale de Lausanne, 
\\CH-1015, Lausanne, Switzerland}


\maketitle
\begin{abstract}
In fusion plasmas the strong magnetic field allows the fast gyro-motion to be systematically removed from the description of the dynamics, resulting in a considerable model simplification and gain of computational time. Nowadays, the gyrokinetic (GK) codes play a major role in the understanding of the development and the saturation of turbulence and in the prediction of the subsequent transport. Naturally, these codes require thorough verification and validation.

Here we present a new and generic theoretical framework and specific numerical applications to test the faithfulness of the implemented models to theory and to verify the domain of applicability of existing GK codes. For a sound verification process, the underlying theoretical GK model and the numerical scheme must be considered at the same time, which has rarely been done and therefore makes this approach pioneering. At the analytical level, the main novelty consists in using advanced mathematical tools such as variational formulation of dynamics for systematization of basic GK code's equations to access the limits of their applicability. The verification of numerical scheme is proposed via the benchmark effort.

In this work, specific examples of code verification are presented for two GK codes: the multi-species electromagnetic ORB5 (PIC) and the radially global version of GENE (Eulerian). 
The proposed methodology can be applied to any existing GK code. We establish a hierarchy of reduced GK Vlasov-Maxwell equations implemented in the ORB5 and GENE codes using the Lagrangian variational formulation.
At the computational level, detailed verifications of global electromagnetic test cases developed from the CYCLONE Base Case are considered, including a parametric $\beta$-scan covering the transition from ITG to KBM and the spectral properties at the nominal $\beta$ value. 
\end{abstract}
\section{Introduction}
The gyrokinetic theory represents one of the most important theoretical frameworks for theoretical and numerical modeling of magnetised plasmas. Historically, it has been considered in the context of fusion plasmas as an accurate tool for assessment of turbulent transport, which in its turn, represents a serious issue for plasma confinement.

At the same time, more recently, gyrokinetic simulations have also been applied to astrophysics \cite{Pueschel_2014}, \cite{Told_PRL_2015} in order to access turbulence at the large spectra of scales, and especially at the small scales, where the MHD approximations fail.

Different numerical implementations of the gyrokinetic equations have been actively developed during the last decades. Since the pioneering implementation of the gyrokinetic Vlasov-Poisson system realised by W.~W. Lee in $1983$ \cite{Lee_1983} in the framework of the Particle-In-Cell method, that approach has undergone important developments and is very popular nowadays, see for example \cite{Hatzky_2007}, \cite{Jolliet_2007}. At the same time, the Eulerian numerical realisation of the gyrokinetic Vlasov-Maxwell equations, which appeared in $1995$ \cite{Kotschenreuther_1995} attracted a similar popularity, for example \cite{Jenko_2000}, \cite{Goerler_2011}. For a detailed review on gyrokinetic simulation see for example \cite{Garbet_Idomura_2010}.

Access to High Performance Computing (HPC) facilities  allowed the theoretical plasma physicists to bring gyrokinetic codes to a significantly more advanced level of realism.
However, the question about the validity of those advanced numerical tools employed for the investigation of new physics has not been sufficiently investigated yet. 

Therefore, a two fold verification framework, which allows one to simultaneously verify the implemented model together with the numerical scheme for the gyrokinetic codes needs to be established. For this purpose, one should deal with two main types of difficulties.
The first difficulty consists in the understanding of the gyrokinetic models implemented in a given gyrokinetic code. This is mainly related to use of the different nomenclature and different orderings for the derivation of the equations of the model. The second difficulty comes from the fact that, the implementation typically uses different discretizations and approximations which may further alter the results. 

A systematic derivation, which guarantees the energetic consistency of gyrokinetic models requires advanced mathematical tools such as differential geometry and variational calculus on functional spaces.
In the electrostatic limit, i.e. in the case of the Vlasov-Poisson equations, a systematic theoretical derivation from the first principles of dynamics has been presented in \cite{Bottino_Sonnendruecker} . For the numerical schemes, numerous verification studies can be found -- amongst others, for instance, the cross-code benchmarks in Refs.~\cite{Dimits_2000}, \cite{McMillan_PRL_2010} and \cite{Lapillonne_2010}.

Regarding the full gyrokinetic Vlasov-Maxwell system including electromagnetic fluctuations, several cross-code comparisons exist in the flux tube limit \cite{GEM_GYRO_GS2, flux_tube_GENE_GKW_GS2,GENE_GYRO_EXB_nonlin}. However, detailed analytic comparisons and radially global benchmarks are hardly to be found which has therefore been identified as one of the goals for the European VeriGyro project launched in 2014.
In very recent works  the theoretical foundations for the Particle-In-Cell codes \cite{TBS_2016} and the result of cross-code benchmark for global electromagnetic gyrokinetic codes,\cite{Goerler_Tronko_2016} have been summarized. 

In this work we the present theoretical framework for the systematic derivation of the ORB5 and GENE codes models as two generic representers of the PIC and Eulerian implementations of the gyrokinetic Vlasov-Maxwell equations.
We also briefly present two inter-code benchmark test cases in order to explicitly illustrate the differences between the implemented models.

In Sec.~\ref{sec:GK_reduction} we review the main ideas behind the gyrokinetic dynamical reduction, paying special attention to comparison between the theoretical assumptions and the code implementations, for example, the difference between the theoretical and code orderings is explained.
In Sec.~\ref{sec:ORB_GENE_gyrocenter} we start with a comparison between the reduced particle (gyrocenter) models for the ORB5 and GENE codes followed in Sec.~\ref{sec:GK_Maxwell_Vlasov} by the presentation of Lagrangian variational formulations for the reduced Maxwell-Vlasov models for both codes. Section~\ref{sec:benchmark} contains the intercode benchmark test cases specific for the code models comparison. The conclusions are summarized into the Sec.~\ref{conclusions}.
\section{\label{sec:GK_reduction}Gyrokinetic dynamical reduction}
Gyrokinetic theory is based on the procedure of an asymptotic dynamical reduction for a multi-scaled dynamical system. Such a procedure is aiming at consrtucting a new set of phase space variables in which the dynamics of a system is restricted on a hypersurface, i.e. an invariant of the motion is serving as one of the phase space variables.
To develop such an asymptotic dynamical reduction, first of all one needs to put in evidence the fact that there exists a dynamical invariant in the considered system and then to define orderings assumed during the asymptotic procedure construction. 

The idea of the gyrokinetic dynamical reduction is based on the physical property of the considered system: the presence of  a strong background magnetic field induces a scales of motion separation on the dynamics of the charged particles moving in the superposition of that strong background guide field and some additional fluctuating perturbative electromagnetic fields.

In fact, particle dynamics is decomposed into the fast rotation around the magnetic field lines and slow drift motion. The cyclotron frequency $\Omega=e B/m c$, where $e$ and $m$ are, respectively, the charge and mass of particles, $B$ is the magnetic field amplitude and $c$ is the speed of light, sets the scale of gyromotion.

The gyromotion is described by a fast gyroangle variable $\theta$ to which corresponds to a canonically conjugated slowly varying magnetic moment $\mu$.
At the lowest order:
\begin{equation}
\mu=m v_{\perp}^2/2 B,
\label{eq:mu0}
\end{equation} 
where $v_{\perp}$ is the perpendicular velocity of particles with respect to the magnetic field lines. 
In the case of a constant and uniform background magnetic field, $\mu$ is an exact dynamical invariant. 

The sources for the violation of the magnetic moment invariance can be attributed to two different reasons. First, the spatial variation of background quantities such as the magnetic field non-uniformity and curvature and, second, the presence of electromagnetic fluctuations.
The gyrokinetic dynamical reduction uses the fact that averaged over the long times magnetic moment is still conserved, i.e. $\left\langle\dot{\mu}\right\rangle_t=0$ even for the perturbed system.

 \subsection{\label{subsec:gyrocenter_variables}Gyrocenter phase space variables}
The goal of the gyrokinetic dynamical reduction consists in building up a new set of phase space variables, such that the $\theta$ dependence is completely uncoupled and the magnetic moment $\mu$ has a trivial dynamics, i.e. $\dot{\mu}=0$. Therefore, the reduced particle dynamics is described in the $5$-dimensional phase space with variables $\left(\mathbf{X},p;\mu\right)$, where $\mathbf{X}$ represents the reduced particle position and $p$ is a scalar moment coordinate and the magnetic moment $\mu$ has a trivial dynamics. This change of coordinate is constructed via a perturbative series of near-identity phase space transformations. These transformations are invertible at each step of the perturbative procedure. 
The particle dynamics on the new reduced phase space is derived within the same near-identity phase space transformation procedure, which is performed on the corresponding phase space Lagrangian. We give the expression in Sec.~(\ref{sec:ORB_GENE_gyrocenter}).

The reduced position $\bf X$ has a simple geometrical meaning: it is the instantaneous center of the fast particle's rotation around the magnetic field line. Therefore, from the space coordinate viewpoint the gyrokinetic transformation is a shift between the initial particle coordinate $\bf x$ and the instantaneous center of its rotation $\bf X$. The difference between both positions is the polarization displacement, defined at the lowest order of the dynamical reduction procedure as the Larmor vector of the particle $\bm\rho_0= mc /e\ (2\mu/m B)^{1/2}\widehat{\bm\rho}$, where $\widehat{\bm\rho}$ represents a fast rotating unit vector, perpendicular to the direction of the background magnetic field $\widehat{\mathbf b}$ and explicitly depending on the fast variable $\theta$. In this work, the exact gyrocenter spatial coordinate transformation is not considered, but instead the lowest order polarization displacement $\bm{\rho}_0$ will be taken in account: it will be shown to be sufficient in particular to expose the main differences between the ORB5 and GENE code models. 
Performing numerical simulations on the $5$-dimensional phase space instead of the $6$-dimensional one results in a drastic reduction of the computational costs. 
\subsection{Numerical schemes}
First of all, let us clarify which system of equations from the mathematical point of view we are about to solve.  In the gyrokinetic model, we have an equation of evolution for the particle distribution function $F$ on the $5$-dimensional phase space, i.e. we have to solve the new partial differential equation on the $5$-dimensional phase space $({\mathbf X},p, \mu)$. This can be solved either by advancing $F$ in time on the $5$D grid (Eulerian method) or by advancing in time the corresponding characteristics, which are $5$ non-linearly coupled ODEs (PIC). In what concerns the field equations, we are dealing with solving the $3$D integro-differential equations: Poisson (quasi-neutrality) and Amp\`ere, which are using the information about the distribution function $F$ in order to evaluate charges and currents. 
To summarize: solving the system of the gyrokinetic Vlasov-Maxwell equations is a rather challenging numerical task.
We provide details of the derivation of these equations below.

In this paper we compare the gyrokinetic Vlasov-Maxwell models implemented in two gyrokinetic codes, which are using two different numerical schemes.

On the one hand, we consider the ORB5 code, which is using the Lagrangian approach for solving the gyrokinetic equations of motion. That approach consists in sampling initial positions in phase space (loading of markers), then following marker orbits in $5$D (pushing) and obtaining the source terms (charge and density) for the field equations at the each time step. 
On the other hand, we consider the GENE code, which is using an Eulerian numerical scheme in order to solve the system of the gyrokinetic Vlasov-Maxwell equations. It consists in discretizing the phase space on a fixed grid, and applying finite differences or finite volume schemes for the differential and integral operators. The Eulerian approach is sometimes also called the "Vlasov" approach.

The GENE code has two versions from the geometrical point of view: the \textit{global} and the \textit{local} version.
The \textit{local} version is also called the \textit{flux-tube} code, in which the domain considered is a vicinity of a magnetic field line. The equations of motion are expanded in that vicinity such that all coefficients, which define the density and temperature profiles are constant: $\bm\nabla n=\mathrm{const}$, $\bm\nabla T=\mathrm{const}$ as well as the geometrical coefficients, i.e. the elements of the metric tensor and in particular $q=\mathrm{const}$ with a constant magnetic shear $\hat{s}=\mathrm{const}$ . 

The ORB5 code has only the \textit{global} version. It means that it takes the geometry of the whole plasma domain into account with consistent plasma profiles and gradients, as well as full metric of the background axisymmetric magnetic geometry.

\subsection{Gyrokinetic theoretical ordering}
 From the point of view of the two-step derivation procedure, in which the effects of the magnetic moment non-invariance induced by the background fields non-uniformities are considered separately from those related to the presence of the electromagnetic fluctuating fields, the small parameters can be organised in two groups. In the first group of parameters, i.e. \textit{guiding-center} transformation related parameters, we include those related to the variations of the background related quantities, i.e. $\epsilon_B=\rho_{\mathrm{th}}/L_B$, where $\rho_{\mathrm{th}}$ is the thermal Larmor radius of the particle and $L_B=|\bm\nabla B/B|^{-1}$ sets up the length scale of the background magnetic field variation.
 
The second group of the small parameters is related to the \textit{gyrocenter} coordinate transformation, aiming to reestablish the invariance of magnetic moment $\mu$, destructed by the presence of fluctuating electromagnetic fields. The associated small parameter can be defined as $\epsilon_{\delta}
\sim|\mathbf{B}_1|/B\sim|\mathbf{E}_{1\perp}|/(B v_{\mathrm{th}})\sim\left(k_{\perp}\rho_{\mathrm{th}}\right)\  e\phi_1/T_i\equiv\epsilon_{\perp}e\phi_1/T_i$, where $v_{\mathrm{th}}$ is the thermal velocity, $B$ the amplitude of the background magnetic field, $T_i$ is the ion temperature and $\phi_1$ represents the amplitude of the fluctuating electrostatic potential. The parameter $\epsilon_{\perp}$ allows the distinction between the gyrokinetic theory with $\epsilon_{\perp}\sim \mathcal{O}(1)$ and the drift-kinetic theory with $\epsilon_{\perp}\ll 1$. In particular, the model which is truncated up to the second order in $\epsilon_{\perp}$ is called the \textit{long-wavelength} approximation of the gyrokinetic theory. The \textit{long-wavelength} approximation is rather popular for the numerical implementations. We will come to that model later in Sec.~\ref{sec:ORB_GENE_gyrocenter}.


The parallel fluctuations of the electric field are pushed on the next level of smallness according to the anisotropy of the turbulence $|E_{1\|}|/|E_{1\perp}|\sim |k_{\|}|/|k_{\perp}|\ll 1$.

In addition to that, for parallel fluctuations of the magnetic field it is assumed that $|B_{1\|}|/|B|\sim\beta\epsilon_{\delta}$, where $\beta=(8\pi p_{\mathrm{th}})/B^2$, the ratio between the kinetic thermal pressure $p_{\mathrm{th}}$ and the magnetic pressure $B^2$, which means that they are only considered when the magnetic beta is close to $1$, i.e. $\beta\sim\mathcal{O}(1)$.

In the case of the \textit{maximal} ordering, the effects from the variations of background quantitites and the electromagnetic fields fluctuations should be considered at the same order $\epsilon_B\sim\epsilon_{\delta}$. 
However, it is important to note, that each set of small parameters, related to a specific choice of physical configuration, will define a setup for the derivation of the corresponding gyrokinetic theory. In other words, it is important to emphasize that there is not only one specific set of reduced Vlasov-Maxwell equations, which is defined as the gyrokinetic Vlasov-Maxwell system, but rather different sets of reduced equations, which need to be put inside the common systematic framework.
\subsection{\label{sec:code_orderings} Gyrokinetic code ordering}
The majority of the gyrokinetic models implemented in the codes are derived within the assumption that $\epsilon_B\ll\epsilon_{\delta}$. Typically, all the background gradient corrections are taken into account at the first order, while the contributions from the fluctuating fields are considered at the second order. Such a treatment allows one to eliminate a significant number of terms, (see for example derivations performed within the \textit{maximal} ordering in \cite{Brizard_2013} and \cite{Tronko_Brizard_2015}) and simplified numerical implementation.

 In what concerns the FLR or the  $\epsilon_{\perp}$- ordering, both models: derived in the limit with full FLR corrections as well as the models truncated up to the second order in $\epsilon_{\perp}$ are implemented. Below, we will discuss differences between the long-wavelength approximated Vlasov-Maxwell models and those containing the full FLR corrections. In addition to that, we note that the ORB5 and the radially \textit{global} version of the GENE code considers the perpendicular fluctuations of the magnetic field only i.e. are restricted to the low-beta ordering only, while the \textit{local} version of the GENE code follows the full $\beta$ ordering.
  
\subsection{Gyroaveraging}

Implementing a gyroaveraging operator is an important and challenging task for the gyrokinetic simulations.We do not focus on the numerical details here, but just provide the theoretical definition, which is necessary for the models derivation.

For each function $\psi$, defined in the position $\mathbf{x}=\mathbf{X}+\bm\rho_0$ we define the \textit{gyroaverage} operator as:
\begin{equation}
\left\langle \psi (\mathbf{X}+\bm\rho_0)\right\rangle=\frac{1}{2\pi}\int_0^{2\pi} d\theta\  \psi(\mathbf{X}+\bm\rho_0(\theta)),
\end{equation}
such that each function on the reduced phase space can be decomposed in the \textit{gyroaveraged} and \textit{fluctuating} parts:
\begin{equation}
\psi=\left\langle \psi \right\rangle+\widetilde{\psi}.
\end{equation}


\section{\label{sec:ORB_GENE_gyrocenter}\textrm{GENE} and \textrm{ORB5} gyrocenter models}
The $6$-dimensional reduced phase space Lagrangian represents a central object of the gyrokinetic dynamical reduction:
\begin{equation}
L_p=\left(\frac{e}{c}{\mathbf A}+\left(\frac{e}{c}\epsilon_{\delta}A_{1\|}+ mv_{\|}\right)\widehat{\mathbf b}+
\epsilon_{\delta}\frac{e}{c} \mathbf{A}_{1\perp}\cdot \ m\mathbf{v}_{\perp}\right)\cdot\dot{\mathbf X}+\frac{mc}{e}\mu\dot{\theta}-H.
\label{L_p}
\end{equation}
The last term in the expression above is referred to as the Hamiltonian of the system. The remaining terms represent the symplectic part of the phase space Lagrangian. This phase space Lagrangian contains the fluctuating electromagnetic fields $(\phi_1,A_{1\|},\mathbf{A}_{1\perp})$, which are explicitly time-dependent. The perturbed fields are evaluated at the spatial position ${\mathbf x}={\mathbf X}+\bm\rho_0$, i.e. they possess the fast gyroangle $\theta$ dependency, which is removed according to the near-identity phase space transformation. That coordinate transformation at the code relevant order is presented in \cite{PPCF_2016}, for the procedure at all orders, see for example \cite{Brizard_Hahm}. Here we skip the detailed description of the reduction procedure and rather focus on its conceptual explanation and comparison between the reduced particle models, issued from the various representations of that procedure.

From the conceptual point of view, the fluctuating electric field $\phi_1$, as a scalar field, is included into the Hamiltonian $H$, whereas, for the magnetic field perturbations $A_{1\|}$ and ${\mathbf A}_{1\perp}$ different options are possible. The choice of including the perturbed magnetic fields $A_{1\|}$ and $\mathbf{A}_{1\perp}$ in the symplectic part or in the Hamiltonian defines its dynamical representation. The right choice of the representation is important for the corresponding numerical scheme realisation. Different dynamical representations are possible, the complete list with the corresponding nomenclature, which we are following below, is available in \cite{Brizard_Hahm}.

Consideration of the perpendicular part of the perturbed electromagnetic potential $\mathbf{A}_{1\perp}$ depends on the choice of the ordering. In the case of the low$\beta$ ordering, when parallel perturbations of magnetic field are neglected, the $\mathbf{A}_{1\perp}$ related term does not appear in the Eq.~ (\ref{L_p}).

It is possible to identify the near-identity phase space transformations, which would affect the Hamiltonian part of the phase space Lagrangian only. For realisation of such a transformation, one should always keep the fluctuating parts of all the perturbed electromagnetic potentials in the Hamiltonian part of $L_p$ and leave the symplectic part free of the explicit fast angle $\theta$ dependency. Here we propose two different options for the definition of such a transformation.

The first option is the Hamiltonian representation (adopted for the ORB5 code) which leaves the symplectic part of the phase space Lagrangian Eq.~(\ref{L_p}) completely free of the perturbative electromagnetic potentials, by including them into the Hamiltonian. That manipulation is possible when using  the canonical parallel gyrocenter moment as one of the phase space variables:
\begin{equation}
p_z=m v_{\|}+\frac{e}{c}\epsilon_{\delta}A_{1\|}(\mathbf{X}+\bm\rho_0).
\end{equation}
The second option is to redistribute the fluctuating potentials between the Hamiltonian and the symplectic parts of the phase space Lagrangian, so that it leaves the symplectic part free of the $\theta$-dependency. This corresponds to the parallel-symplectic representation. In that case, only the \textit{gyroaveraged} component of the parallel part of the perturbed magnetic potential $\left\langle A_{1\|}\right\rangle$ is contained in the symplectic part of the phase space Lagrangian and its fluctuating part together with the perpendicular part of the perturbed magnetic potential is accounted in the Hamiltonian. This parallel-symplectic representation is used for the GENE code, however with some further approximations as explained below.

The parallel-symplectic representation, from the theoretical point of view is using a modified parallel moment consisting of the sum of the parallel kinetic moment and the fluctuating part of the parallel magnetic  potential $\bar{p}_{\|}=mv_{\|}+\frac{e}{c}\widetilde{A}_{1\|}(\mathbf{X}+\bm\rho_0)$ of the particle as the phase space variable. However,  the fluctuating component $\widetilde{A}_{1\|}$ is neglected in the GENE code model and therefore the phase space variable is $p_{\|}=mv_{\|}$, i.e. the kinetic moment.

We define the symplectic magnetic potentials in the following form:
\begin{eqnarray}
\mathbf{A}^*&=&\mathbf{A}+\frac{c}{e}p_z\widehat{\mathbf{b}},
\\
\mathbf{A}^{**}&=&\mathbf{A}+\left(\epsilon_{\delta}\left\langle A_{1\|}\right\rangle+\frac{c}{e}p_{\|}\right)\widehat{\mathbf b},
\end{eqnarray}
therefore the reduced particle phase space Lagrangians are
\begin{eqnarray}
L_p^{*}&=&\frac{e}{c}\mathbf{A}^*\cdot\dot{\mathbf X}+\frac{mc}{e}\mu\dot{\theta}-H^*,
\label{eq:L_*}\\
L_p^{**}&=&\frac{e}{c}\mathbf{A}^{**}\cdot\dot{\mathbf X}+\frac{mc}{e}\mu\dot{\theta}-H^{**},
\label{eq:L_**}
\end{eqnarray}
where $H^*$ and $H^{**}$ are the Hamiltonians, corresponding to the hamiltonian or parallel-symplectic representation of the dynamics, respectively.

As we can see, in the first case the symplectic part of the phase space Lagrangian is time independent and in the second case the symplectic part of the phase-space Lagrangian is explicitly time-dependent, because of the presence of $\left\langle A_{1\|}\right\rangle$. That has a direct impact on the equations for the corresponding characteristics of the dynamical variables on the reduced phase space.
The Hamiltonian representation allows one to avoid explicit time derivative of the gyroaveraged parallel magnetic potential $\left\langle A_{1 \|}\right\rangle$ on the r.h.s. of the reduced phase space $(\mathbf{X},p_{\|})$ characteristics, while the parallel-symplectic representation makes it appear explicitly in the characteristic for the kinetic parallel moment $p_{\|}$, see for example \cite{TBS_2016}.


In the table below we summarize the reduced  particle models implemented in both of the codes: first, we explicit the symplectic and then the Hamiltonian parts of the phase-space Lagrangian.
\begin{table}[h!]
\begin{tabular}{c c}
ORB5 (Hamiltonian, low $\beta$) & GENE (Modified parallel symplectic)
\\
&
local: $\alpha=1$; global $\alpha=0$
\\
\hline\\
$\mathbf{A}^*=\mathbf{A}+ (c/e)\ p_z\widehat{\mathbf b}$&
$\mathbf{A}^{**}=\mathbf{A}+ (c/e)\ {p}_{\|}\widehat{\mathbf b}+\epsilon_{\delta}\left\langle A_{1\|}(\mathbf{X}+\bm\rho_0)\right\rangle\widehat{\mathbf b}$
\\
\newline
\\
$
H_0^{\mathrm{ORB5}}=p_{z}^2/(2m)+\mu B
$
&
$
H_0^{\mathrm{GENE}}=p_{\|}^2/(2m)+\mu B
$
\\
$
\newline
$
\\
$H_1^{\mathrm{ORB5}}=e\left\langle\psi_1^{\mathrm{ORB5}}\right\rangle$,
& 
$H_1^{\mathrm{GENE}}=e\left\langle\psi_1^{\mathrm{GENE}}\right\rangle$,
\\
with
$
\psi_1^{\mathrm{ORB5}}=\phi_1-A_{1\|}p_{z}/m
$
&
with
$
\psi_1^{\mathrm{GENE}}=\phi_1-\alpha{\mathbf A}_{1\perp}\cdot\mathbf{v}_{\perp}
$
\\
\newline
\\
$
H_2^{\mathrm{ORB5}}=-\frac{e^2}{2B}\frac{\partial}{\partial\mu}\left\langle\widetilde{\phi}_1\left({\mathbf{X}+\bm\rho_0}\right)^2\right\rangle
$
&
$
H_2^{\mathrm{GENE}}=-\frac{e^2}{2B}\frac{\partial}{\partial\mu}\left\langle\widetilde{\psi}_1^{\mathrm{GENE}}(\mathbf{X}+\bm\rho_0)^2\right\rangle
$
\\
$
+\frac{e^2}{2 mc^2}\left(A_{1{\|}}(\mathbf{X})^2+m\left(\frac{c}{e}\right)^2 \frac{\mu}{B}A_{1\|}(\mathbf{X})\bm\nabla_{\perp}^2 A_{1\|}\left(\mathbf{X}\right)\right)
$
&
\\
\newline
\\
$
H_{2,\mathrm{FLR}}^{\mathrm{ORB5}}=-\frac{m c^2}{2B^2}\left|\bm\nabla\phi_1\left({\mathbf{X}}\right)\right|^2
$
&
$
$
\\
$
+\frac{e^2}{2 mc^2}\left(A_{1{\|}}(\mathbf{X})^2+m\left(\frac{c}{e}\right)^2 \frac{\mu}{B}A_{1\|}\bm\nabla_{\perp}^2 A_{1\|}\left(\mathbf{X}\right)\right)
$
&
\end{tabular}
\caption{\label{tab:Hamiltonians}Reduced Hamiltonian dynamics comparison.}
\end{table}
In table I, the modified and parallel-symplectic model means that the fluctuating component of the magnetic potential $\widetilde{A}_{1\|}$ has been ignored in the definition of the phase space variable $\bar{p}_{\|}$, which in its turn leads to neglect the $\left\langle\widetilde{A}_{1\|}(\mathbf{X}+\bm\rho_0)^2\right\rangle$ term into the expression for the second order Hamiltonian; additionally the square of the perpendicular component of the electromagnetic potential $A_{1\perp}\left({\mathbf X}+\bm\rho_0\right)^2$ has also been omitted.
\section{\label{sec:GK_Maxwell_Vlasov}Gyrokinetic Vlasov-Maxwell equations}
There exist two significantly different approaches for the derivation of the reduced gyrokinetic Vlasov-Maxwell equations. First of all, one can proceed with the direct calculation of the moments for the reduced gyrokinetic Vlasov distribution function in order to evaluate the charge and the current densities in the gyrokinetic Poisson and Amp\`ere equations (zeroth order moment corresponding to gyrokinetic Poisson and the first moment to gyrokinetic Amp\`ere equation). This approach is called the "pull-back" transformation.  Another possible approach is to get the reduced gyrokinetic field equations from the variational formulation in which the interaction between the reduced particle dynamics, described by the particle Lagrangian $L_p$ and the dynamics of the electromagnetic fields is included inside the field-particle Lagrangian. 

The first approach is an intuitive approach, it has been introduced, for instance, in Ref.~\cite{Hahm_1988} within the electrostatic approximation. On the other hand, the variational approach is more formal and it has been formulated almost two decades later in Refs.~\cite{Sugama_2000} and \cite{brizard_prl_2000}.
For historical reasons, the GENE code equations have been derived within this intuitive approach, while the ORB5 model is already obtained within the variational framework.

To make a formal comparison between the models of both codes in this work we choose to derive the theoretical code models from the variational formulation of the reduced dynamics \cite{Sugama_2000}.

\subsection{Gyrokinetic field formulation}
In this section we present the variational framework for the derivation of consistently coupled gyrokinetic Vlasov-Maxwell equations suitable for a code implementation. 
The first term of the field-particle Lagrangian includes the reduced particle dynamics represented by the gyrocenter Lagrangian $L_p$ coupled to the Vlasov distribution function $F$; the second term contains the electromagnetic fields:

\begin{eqnarray}
\mathcal{L}&=&\sum_{\mathrm{sp}} \int dV dW F(\mathbf{Z}_0,t_0) L_p\left(\mathbf{Z}\left[\mathbf{Z}_0,t_0;t\right],\dot{\mathbf{Z}}\left[\mathbf{Z}_0,t_0;t\right];t
\right)
+\int dV \ \frac{\left|\mathbf{E}_1\right|^2-\left|\mathbf{B}_1\right|^2}{8\pi},
\label{F_P_Lagrangian}
\end{eqnarray}
where the reduced phase space variables are $\mathbf{Z}=\left({\mathbf X}, p, \mu,\theta \right)$ with $d{\mathbf X}=d V$ and the reduced velocity phase space volume is chosen according to the representation of the reduced particle dynamics $dW=B_{\|}^{\#}dp\ d\mu$, i.e. $B_{\|}^{\#}=(B^*_{\|},B^{**}_{\|})$ and $p=(p_z,p_{\|})$.

The main important idea about building up a consistent gyrokinetic Vlasov-Maxwell model implementable into the given code consists in the fact that \textit{all} the approximations should be performed on the field-particle Lagrangian \textit{before} the derivation of the gyrokinetic Vlasov-Maxwell equations. The corresponding equations of motion should be obtained according to the first principle of dynamics together with the corresponding conservation laws following the Noether method.

%
\subsection{ORB5 and GENE codes models derivation}
%
Here, we focus on the derivation of the ORB5 and GENE \textit{theoretical} models.
 By \textit{theoretical} we mean that the derived equations are written down in the form, which follows directly from the analytical calculation. Note that each numerical scheme has its own requirement for model rewriting before discretization, aiming to simplify the numerical resolution.
 We do not focus on the detailed comparison of the discretized models here. In order to perform the verification of the numerical implementations we realise a detailed intercode benchmark, linear results can be found in \cite{Goerler_Tronko_2016}. Two test cases connected to the \textit{theoretical} models verification are presented in Sec.~\ref{sec:benchmark}.

We provide a detailed list of approximations performed on the field particles Lagrangian (\ref{F_P_Lagrangian}) in order to obtain the models corresponding to the ORB5 and GENE codes. We start with presenting the approximations common to both codes and and subsequently focus on the specific details of each model.

The first common approximation considering the derivation of the ORB5 and GENE code models consists in the fact that the field-particle Lagrangian (\ref{F_P_Lagrangian}) is truncated up to the second order i.e. contains up to the $\mathcal{O}(\epsilon_{\delta}^2)$ electromagnetic field terms. It means that the second order Lagrangian couples nonlinear terms related to the reduced particle dynamics $H_2$, to the background (non-dynamical) Vlasov distribution function $F_0$ only.
Therefore, the corresponding gyrokinetic Vlasov equation contains exclusively the linear (i.e. $\sim\mathcal{O}(\epsilon_{\delta})$) terms.

The second common approximation, directly applied to the field particle Lagrangian (\ref{F_P_Lagrangian}) is the \textit{quasi-neutrality} approximation. We compare the field term with the nonlinear second order particle contribution.
According to the Tab.~\ref{tab:Hamiltonians} both Hamiltonian second order reduced particle models $H_2^{\mathrm{ORB5}}$ and $H_2^{\mathrm{GENE}}$ coincide in the electrostatic limit . Therefore, in the \textit{long-wavelength} approximation $H_{2,\mathrm{FLR}}^{\mathrm{es}}=-mc^2/2B^2|\bm\nabla_{\perp}\phi_1|^2$. With taking into account the gyrokinetic ordering for the electrostatic field with $E_{1\|}/E_{1\perp}=\epsilon_{\delta}$, we neglect the parallel contribution $E_{1\|}=\nabla_{||}\phi_1-1/c\ \partial_t A_{1\|}$ to the electric field $\mathbf{E}_1$ in the second term of the Eq.~(\ref{F_P_Lagrangian}).
Therefore,
\begin{eqnarray}
&&\int dV\  \frac{|\mathbf{E}_1|^2}{8\pi}+\int dW\ dV\ F_0\ \frac{mc^2}{2 B^2}|\bm\nabla_{\perp}\phi_1|^2
\label{quasi}
\\
&=&
\alpha \int dV\  \frac{1}{c^2}\left|\frac{\partial\mathbf A_{\perp1}}{\partial t}\right|^2+\frac{1}{8\pi}\int dV\ \left(1+\frac{\rho_s^2}{\lambda_d^2}\right)|\bm\nabla_{\perp}\phi_1|^2,
\nonumber
\end{eqnarray}
$\alpha=0$ corresponds to the low-beta approximation of the reduced dynamics and $\alpha=1$, to the finite $\beta$ approximation. In the strongly magnetised plasma the ratio of the sound Larmor radius and the Debye length $\rho_s^2/\lambda_d^2=c^2/v_A^2\gg 1$, i.e. we can systematically neglect the term $1/8\pi \int dV \left|\bm\nabla_{\perp}\phi_1\right|^2$ in Eq.~(\ref{quasi}).

 The last common code model approximation is performed on the perturbed part of the magnetic field, i.e. 
\begin{equation}
\mathbf{B}_{1 \perp}=\widehat{\mathbf{b}}\times\bm\nabla A_{1\|},
\label{magnetic}
\end{equation}
which means that we have neglected the  $A_{1\|}\bm\nabla\times\widehat{\mathbf b}\sim\mathcal{O}(\epsilon_{\delta}\epsilon_B)$ term according to the general codes ordering, which we have discussed in Sec.~\ref{sec:code_orderings}.
\subsection{Gyrokinetic Maxwell equations for ORB5}
The second order \textit{linearised} field-particle Lagrangian used in ORB5, including the second order Hamiltonian $H_2^{\mathrm{ORB5}}$ as written in Tab.~\ref{tab:Hamiltonians} reads:
\begin{eqnarray}
\mathcal{A}^{\mathrm{ORB5}}=\int_{t_0}^{t_1}\ dt\ \mathcal{L}^{\mathrm{ORB5}}&=&\sum_{\mathrm{sp}} \int dt\ dV dW
\left(\frac{e}{c}\mathbf{A}^*\cdot\dot{\mathbf{X}}+\frac{mc}{e}\mu\dot{\theta}-\left(H_0^{\mathrm{ORB5}}+\epsilon_{\delta} H_1^{\mathrm{ORB5}}\right)
\right)\left(F_0+\epsilon_{\delta}F_1\right)
\nonumber
\\
&-&\epsilon_{\delta}^2\sum_{\mathrm{sp}} \int\ dt\  dV dW \ H_{2}^{\mathrm{ORB5}} F_0
-\epsilon_{\delta}^2\int\ dt\ dV \ \frac{\left|\mathbf{B}_{1 \perp}\right|^2}{8\pi}
\label{L_ORB5}
\end{eqnarray}
where we have assumed the approximation on the magnetic field given by Eq.~(\ref{magnetic}).
With using the first principle of dynamics, we derive the gyrokinetic Vlasov-Maxwell equations corresponding to field-particle Lagrangian given by Eq.~(\ref{L_ORB5}). Here we limit our derivation to the weak form of the equations of motion, it means, including the arbitrary test function $\widehat{A}_{1\|}$ and $\widehat{\phi}_{1}$. The weak form is suitable for the finite element discretisation implemented in the ORB5 code. 
We start our derivation with the gyrokinetic quasineutrality equation:
\begin{eqnarray}
0=\frac{\delta\mathcal{L}^{\mathrm{ORB5}}}{\delta\phi_1}\circ\widehat{\phi}_1
\Rightarrow 0&=&
-\epsilon_{\delta}\int dV\ dW\ \left(\frac{e^2}{B} F_0\right) \partial_{\mu}\left\langle\widetilde{\phi}_1\left(\mathbf{X}+\bm\rho_0\right)\widehat{\phi}_1(\mathbf{X}+\bm\rho_0)\right\rangle
\nonumber
\\
&+&e\int\ dV\ dW \left(F_0+\epsilon_{\delta}F_1\right)\left\langle\widehat{\phi}_1(\mathbf{X}+\bm\rho_0)\right\rangle.
\end{eqnarray}
The Amp\`ere equation is:
\begin{eqnarray}
0&=&\frac{\delta\mathcal L^{\mathrm{ORB5}}}{\delta A_{1\|}}\circ{\widehat{A}_{1\|}}
=
\\
\nonumber
&-&\epsilon_{\delta}\int dV\ \frac{1}{4\pi}\left[
\bm\nabla\times \left({A}_{1\|}\widehat{\mathbf{b}}\right)\right]\cdot\left[\bm\nabla\times\left(\widehat{A}_{1\|}\widehat{\mathbf{b}}\right)
\right]
\nonumber
+
\int dV\ dW\ \left(F_0+\epsilon_{\delta}F_1\right) \left\langle p_z\widehat{A}_{1\|}(\mathbf{X}+\bm\rho_0)\right\rangle
\\
&-&\epsilon_{\delta}\frac{e^2}{mc^2}\int dV\ dW\ F_0\left( A_{1\|}(\mathbf{X}) \widehat{A}_{1\|}(\mathbf{X})+
m\left(\frac{c}{e}\right)^2 \frac{\mu}{B}\left(A_{1\|}(\mathbf{X})\bm\nabla_{\perp}^2\widehat{A}_{1\|}\left(\mathbf{X}\right)+\widehat{A}_{1\|}\left(\mathbf{X}\right)\bm\nabla_{\perp}^2 A_{1\|}\left(\mathbf{X}\right)\right)\right).
\nonumber
\end{eqnarray}

\subsection{Gyrokinetic Maxwell equations for GENE}
The second order (i.e. containing terms up to $\mathcal{O}(\epsilon_{\delta}^2)$) linearised field-particle Lagrangian action with the second order Hamiltonian  $H_2^{\mathrm{GENE}}$ for the GENE code is given by:

\begin{eqnarray}
\mathcal{A}^{\mathrm{GENE}}=\int_{t_0}^{t_1}\ dt\ \mathcal{L}^{\mathrm{GENE}}&=&\sum_{\mathrm{sp}} \int\ dt\ dV dW
\left(\frac{e}{c}\mathbf{A}^{**}\cdot\dot{\mathbf{X}}+\frac{mc}{e}\mu\dot{\theta}-\left(H_0^{\mathrm{GENE}}+\epsilon_{\delta} H_1^{\mathrm{GENE}}\right)
\right)\left(F_0+\epsilon_{\delta} F_1\right)
\nonumber
\\
&-&\epsilon_{\delta}^2\sum_{\mathrm{sp}} \int\ dt\ dV dW \ H_{2}^{\mathrm{GENE}} F_0
-\epsilon_{\delta}^2\int dV \ \frac{\left|\mathbf{B}_{1}\right|^2}{8\pi},
\label{eq:A_GENE}
\end{eqnarray}
where we have taken into account the fact that the term $A_{1\|}\bm\nabla\times\widehat{\mathbf b}\sim\mathcal{O}(\epsilon_{\delta}\epsilon_B)$ is  neglected, so that only the part of $\mathbf{B}_{1\perp}$ given by Eq.~(\ref{magnetic}) is taken into account, such that ${\mathbf B}_1=\alpha \bm\nabla\times{\mathbf A}_{1\perp}+\widehat{\mathbf b}\times\bm\nabla A_{1\|}$, with $\alpha=0$ for the global and $\alpha=1$ for the local (\textit{flux-tube}) code.
This approximation further affects the field-particle Lagrangian given by Eq.~(\ref{eq:A_GENE}): the symplectic magnetic field $\mathbf{B}^{**}=\bm\nabla\times\mathbf{A}^{**}$ is approximated up to:
\begin{equation}
\mathbf{B}^{**}\approx {\mathbf B}+\frac{c}{e}p_{\|}\bm\nabla\times\widehat{\mathbf b} -\widehat{\mathbf b}\times\bm\nabla\left\langle A_{1\|}\right\rangle,
\label{B_star_approx}
\end{equation}
which means that  the parallel component of the symplectic magnetic field coincides with the one from the Hamiltonian representation of dynamics, i.e.
\begin{equation}
B_{\|}^{**}=\widehat{\mathbf{b}}\cdot{\mathbf B}^{**}\approx B_{\|}^*.
\label{B_par_approx}
\end{equation}
In addition to that the volume element of the velocity phase space $dW$ is further approximated up to $dW=B\ d p_{\|}\ d\mu\ d\theta$. 

Following the same procedure as for the ORB5 code, we derive the corresponding gyrokinetic Vlasov-Maxwell equations from the first principle of dynamics in the weak form. The weak form is considered for the purpose of comparison of the models implemented in ORB5 and GENE codes in the same form.

We start with the gyrokinetic quasineutrality equation:
\begin{eqnarray}
0=\frac{\delta\mathcal L^{\mathrm{GENE}}}{\delta\phi_1}\circ\widehat{\phi}_1&=&
\int dV\ dW\ \left(F_0+\epsilon_{\delta}F_1\right) \left\langle\widehat{\phi}_1(\mathbf{X}+\bm\rho_0)\right\rangle
\\
\nonumber
&+&\epsilon_{\delta}
\int dV\ dW F_0\ \frac{1}{B}\left\langle\partial_{\mu}\left(e\widetilde{\phi}_1\left(\mathbf{X}+\bm\rho_0\right)-\alpha\ \widetilde{\mathbf{v}_{\perp}\cdot\mathbf{A}_{1\perp}}\left(\mathbf{X}+\bm\rho_0\right)\right) \widehat{\phi}_1(\mathbf{X}+\bm\rho_0)\right\rangle.
\end{eqnarray}
The parallel component of the gyrokinetic Amp\`ere equation:
\begin{eqnarray}
0=\frac{\partial\mathcal L^{\mathrm{GENE}}}{\partial A_{1\|}}\circ{\widehat{A}_{1\|}}
&=-&\epsilon_{\delta}\int dV\ \frac{1}{4\pi}\left[
\bm\nabla\times \left({A}_{1\|}\widehat{\mathbf{b}}\right)\right]\cdot\left[\bm\nabla\times\left(\widehat{A}_{1\|}\widehat{\mathbf{b}}\right)
\right]
\\
\nonumber
&+&
\int dV\ dW \ \left(F_0+\epsilon_{\delta}F_1\right) \left\langle p_{\|}\widehat{A}_{1\|}(\mathbf{X}+\bm\rho_0)\right\rangle.
\end{eqnarray}
Finally, for the perpendicular component of the gyrokinetic Amp\`ere equation (for the local GENE code only, i.e. with $\alpha=1$):
\begin{eqnarray}
0&=&\frac{\delta\mathcal L^{\mathrm{GENE}}}{\delta \mathbf{A}_{1\perp}}\circ{\widehat{\mathbf{A}}_{1\perp}}
=-\epsilon_{\delta}\int dV\ \frac{1}{4\pi}\left(\bm\nabla\times\mathbf{A}_{1\perp}\right)\cdot\left(\bm\nabla\times\widehat{\mathbf A}_{1\perp}\right)+\int dV\ dW \left(F_0+\epsilon_{\delta} F_1\right) \left\langle\mathbf{v}_{\perp}\cdot\widehat{\mathbf A}_{1\perp}(\mathbf{X}+\bm\rho_0)\right\rangle
\nonumber\\
&-&
\epsilon_{\delta}\ \int dV\ dW F_0\ \frac{1}{B}\left\langle\partial_{\mu}\left(e\widetilde{\phi}_1\left({\mathbf X}+\bm\rho_0\right)-
\alpha\ \widetilde{\mathbf{v}_{\perp}\cdot{\mathbf{A}}_{1\perp}}\left({\mathbf X}+\bm\rho_0\right)\right)\mathbf{v}_{\perp}\cdot\widehat{\mathbf A}_{1\perp}(\mathbf{X}+\bm\rho_0)\right\rangle.
\end{eqnarray}
\subsection{Discussion}
Let us now consider the main differences between the gyrokinetic Maxwell equations for both codes.
The quasineutrality and the linear parallel Amp\`ere equations for the ORB5 code and the ones for the \textit{global} version of the GENE code differs only with respect to the phase space volume element: ORB5 code uses $B_{\|}^*$ and GENE implementation uses $B$. 
Terms proportional to the perturbed parallel magnetic potential  $A_{1\|}$  appearing in the last term of the parallel Amp\`ere equation are available in the ORB5 code up to the second order FLR decomposition. We note that those terms are not present in the GENE code.
Finally, the ORB5 code and the \textit{global} GENE code do not consider the perpendicular Amp\`ere equation yet.
%
\subsection{Gyrokinetic Vlasov equations for ORB5 and GENE codes}
%
The gyrokinetic Vlasov equations  for the ORB5 and GENE codes are reconstructed from the characteristics (i.e. dynamical equations for the phase space variables $\mathbf X$ and $p$). The characteristics, in their turn, are obtained from the fields-particles Lagrangian action, corresponding to the codes given by Eqs.~(\ref{L_ORB5}) and (\ref{eq:A_GENE}).
\begin{eqnarray}
\frac{\delta{\mathcal A}}{\delta{\mathbf Z}}=0\Rightarrow \frac{\delta{\mathcal L}}{\delta{\mathbf Z}}=0
\end{eqnarray}
and therefore, the Euler-Lagrange equations for the characteristics are given by:
\begin{eqnarray}
\frac{d}{dt}\frac{\partial L_p}{\partial \dot{\mathbf Z}}=\frac{\partial L_p}{\partial{\mathbf Z}}.
\label{Lagrangian_characteristics}
\end{eqnarray}
Both codes are using the first order in $\epsilon_{\delta}$, linearized characteristics, i.e. corresponding to the linearized Hamiltonian 
 \begin{equation}
 H=H_0+\epsilon_{\delta}H_1,
 \label{H_linear}
 \end{equation}
where $H_0$ and $H_1$ are chosen correspondingly to the code model from the Tab.~\ref{tab:Hamiltonians}.
 
In the case of the ORB5 code, which corresponds to the Hamiltonian representation of dynamics, we have:
\begin{eqnarray}
\dot{\mathbf X}&=&\frac{c\widehat{\mathbf b}}
{e B_{\|}^{*}}\times\bm\nabla H^{\mathrm{ORB5}}+\frac{\partial H^{\mathrm{ORB5}}}{\partial{{p}_{z}}}\frac{\mathbf{B}^{*}}{B_{\|}^{*}}\\
\dot{p}_{z}&=&-\frac{\mathbf{B}^{*}}{B_{\|}^{*}}\cdot \bm\nabla H^{\mathrm{ORB5}}. 
\nonumber
\end{eqnarray}
The detailed derivation of the equations of motion for the ORB5 code in the case of $H_{2\ \mathrm{FLR}}^{\mathrm{ORB5}}$ and their comparison with the theoretical model, containing up to the second order terms in $\epsilon_{\delta}$ can be found in \cite{TBS_2016}.

While in the case of the GENE code, which corresponds to the modified parallel-symplectic representation of dynamics, we have:
\begin{eqnarray}
\dot{\mathbf X}&=&\frac{c\widehat{\mathbf b}}
{e B_{\|}^{**}}\times\bm\nabla H^{\mathrm{GENE}}+\frac{\partial H^{\mathrm{GENE}}}{\partial{{p}_{\|}}}\frac{\mathbf{B}^{**}}{B_{\|}^{**}}\\
\dot{{p}}_{\|}&=&-\frac{\mathbf{B}^{**}}{B_{\|}^{**}}\cdot\left(\bm\nabla H^{\mathrm{GENE}} +\frac{e}{c}\frac{\partial}{\partial t}\left\langle A_{1\|}\left({\mathbf X}+{\bm\rho}_0\right)\right\rangle\widehat{\mathbf b}\right),
\nonumber
\end{eqnarray}
where the numerical implementation into GENE takes into account the approximation on the symplectic magnetic field $\mathbf{B}^{**}$ given by Eqs.~(\ref{B_star_approx}) and (\ref{B_par_approx}).

For both of the codes, the gyrokinetic Vlasov equation is reconstructed from the characteristics in the following way:
\begin{eqnarray}
0=\frac{d (F_0+\epsilon_{\delta}F_1)}{dt}=\frac{\partial (F_0+\epsilon_{\delta}F_1)}{\partial t}
+\dot{\mathbf X}\cdot\bm\nabla (F_0+\epsilon_{\delta}F_1)+\dot{p}\frac{\partial(F_0+\epsilon_{\delta}F_1)}{\partial p},
\label{eq:Vlasov_general}
\end{eqnarray}
where $p=(p_z,p_{\|})$ depending on the code. 

Let us now analyze the terms in the Vlasov equation. First of all, both codes take into account that the background distribution function is time independent, i.e. $d F_0/d t=0$.

In what concerns the ORB5 code all the nonlinear terms are implemented and no approximation has been made for the symplectic magnetic field $\mathbf{B}^*$, which corresponds to the Hamiltonian representation. The latter guarantees that the phase space volume is preserved following the gyrokinetic coordinate transformation.
For the GENE code, however, the symplectic magnetic field is approximated up to the term $\left\langle A_{1\|}\right\rangle\bm\nabla\times\widehat{\mathbf{b}}\sim\mathcal{O}(\epsilon_B\epsilon_{\delta})$, an additional ordering is implemented, which allows  one to eliminate the $\partial_{p_{\|}}$-derivatives in non-linear terms of the Vlasov equation, given by Eq.~(\ref{eq:Vlasov_general}) is implemented.
\section{\label{sec:benchmark}Numerical verification: ORB5/GENE Benchmark}
Establishing a framework for the detailed comparison
of the basic equations naturally represents just one
pillar in a code comparison. The second is to verify
the correct numerical implementation of the underlying
model by either comparisons with analytic results or
via benchmarks with codes based on entirely different
numerical approaches.
In this section, we present examples for the latter
which have been realised within the framework of
the EUROfusion project VeriGyro as well. The detailed
setups and results involving up to five different codes
can be found in \cite{Goerler_Tronko_2016}. Here, however, we focus on
the results from the PIC based ORB5 and the Eulerian
code GENE in order to complement the theoretical
model comparison introduced in the previous sections.
In particular, we will focus on two specific test cases.
The first one aims at cross-checking the stabilization
of ion temperature gradient (ITG) driven modes
with $\beta$ and the eventual onset of kinetic ballooning
modes (KBMs) above a certain threshold value. Here, we
employ linear growth rates and real frequencies as
observables.
Since comparisons of linear modes are not limited to these
two quantities, a second test case at a given toroidal
mode number $n$ and for a fixed finite $\beta$ is presented.
Here, full poloidal and radial profiles of the electrostatic
$\phi_1$ and the parallel electromagnetic $A_{1\|}$ potentials are
considered. These extended comparisons furthermore reveal
and illustrate the impact of including the full FLR corrections
into the model for the reduced particles dynamics.
The latter is achieved by comparing the full FLR
solver for the gyrokinetic Poisson equation implemented
in GENE code with the long-wavelength approximation
solver implemented in the ORB5 code version, which has
been used during the benchmark campaign.

\subsection{Electromagnetic $\beta$ scan at fixed wave number}

In Fig. 1, we present the result of the electromag-
netic $\beta$-scan at the fixed toroidal wave number $n = 19$.
As mentioned in Sec.\ref{sec:code_orderings} both ORB5 and GENE global
codes are using the same low $\beta$ ordering (both codes
neglect the $A_{1\perp}$ part of the electromagnetic potential).
As one can see, the linear growth rates and
real frequencies from both codes - marked by red and blue
lines - demonstrate good agreement. The plot furthermore
contains the results from the local (flux-tube) GENE version
maximized over radius and ballooning angle (black lines) in
order to quantify finite size effects and emphasize the
need for global electromagnetic models in the scenario at
hand. The flux tube growth rates are indeed found to be
generally higher. More strikingly, the mode transition is
observed at a different $\beta$ value since the finite size
(so-called $\rho^*$) effects seem to be depend on the
microinstability type as well.
Differences between the local (flux-tube) and
global results have been found in previous linear comparisons
without electromagnetic effects and have furthermore been
confirmed in full electrostatic $\beta = 0$ (nonlinear)
turbulence simulations, see e.g. \cite{McMillan_PRL_2010}
and references therein.
For the case shown here, we have $\rho^*\approx 1/182$ such that
mild but visible deviations as found in Fig. 1 are indeed
expected.
\begin{figure}[h!]
\centering
\begin{tabular}{c}
\includegraphics[height=4cm,width=0.7\linewidth]{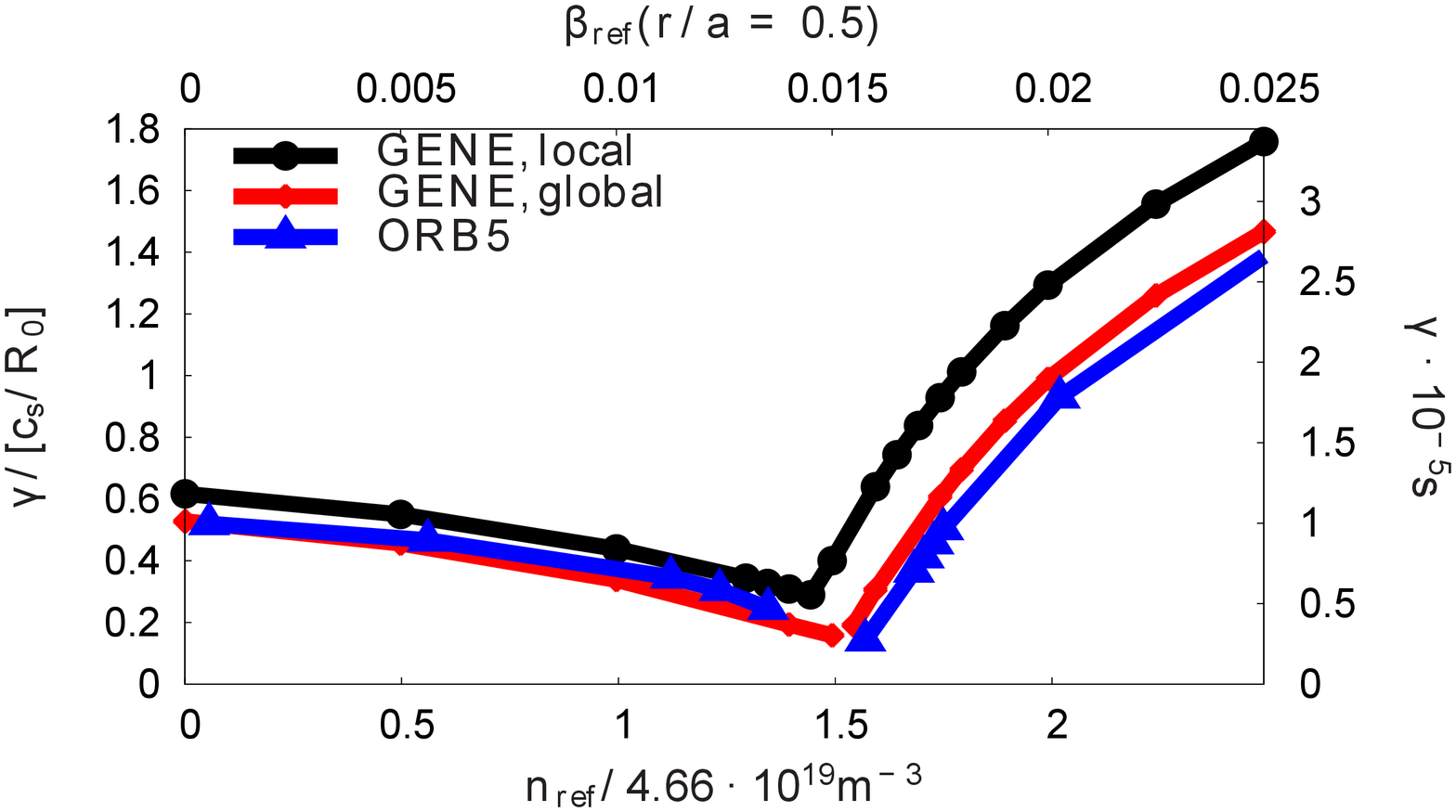}
\\
\includegraphics[height=4cm,width=0.7\linewidth]{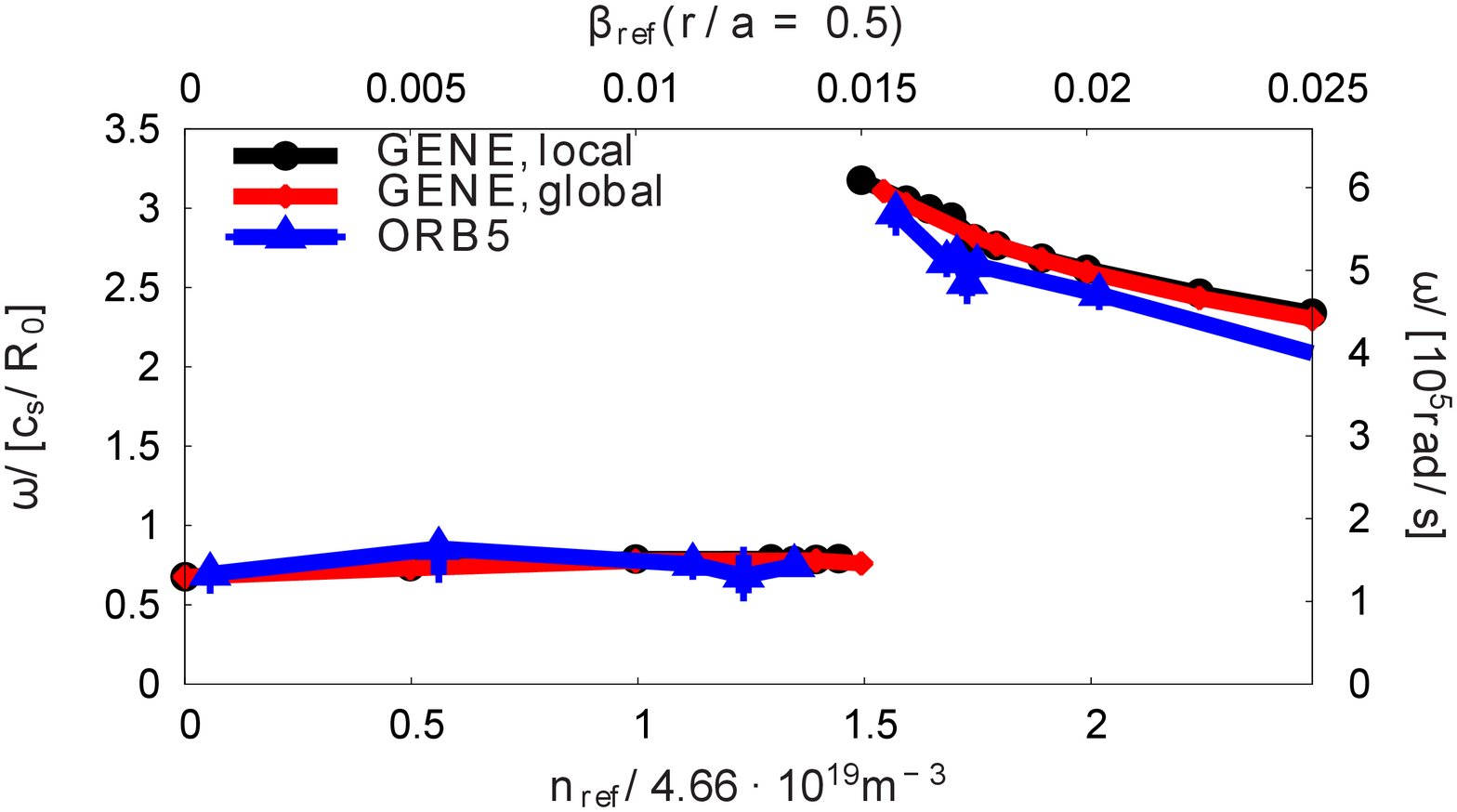}
\end{tabular}
\caption{GENE/ORB5 beta scan. Toroidal wave number $n=19$.}
\label{beta_scan_elm}
\end{figure}
\subsection{Radial and poloidal mode structures at fixed $\beta$}

Another example for a direct comparison of linear
electromagnetic microinstabilities is found in Fig.~2.
Here we compare radial and poloidal mode structures of
the electrostatic and magnetic potential.
While generally agreement between the two codes at hand is
confirmed, they particularly differ with regard to
fine-scale structures in the radial profile of the
electrostatic potential which
can be related to mode rational surfaces and which are
absent in the ORB5 profiles. The reason is the choice
of a long-wavelength approximation Poisson solver --
the full FLR solver, which has
been recently implemented in ORB5 \cite{Dominski_2017}, should recover
the peaks. Considering the good agreement in Fig.~1,
missing the radial fine-scale structures does not introduce
too much harm to visibly alter the growth rates and
frequencies. However, as discussed in \cite{Goerler_Tronko_2016} in more
detail, deviations will become intolerable at the latest
if higher toroidal mode numbers $n\gtrsim 40$ are considered.
This problem is also remedied by the new solver option in
ORB5.
\begin{figure}
\centering
\begin{tabular}{c c}
\begin{minipage}{.25\textwidth}
\includegraphics[width=\linewidth, height=3cm]{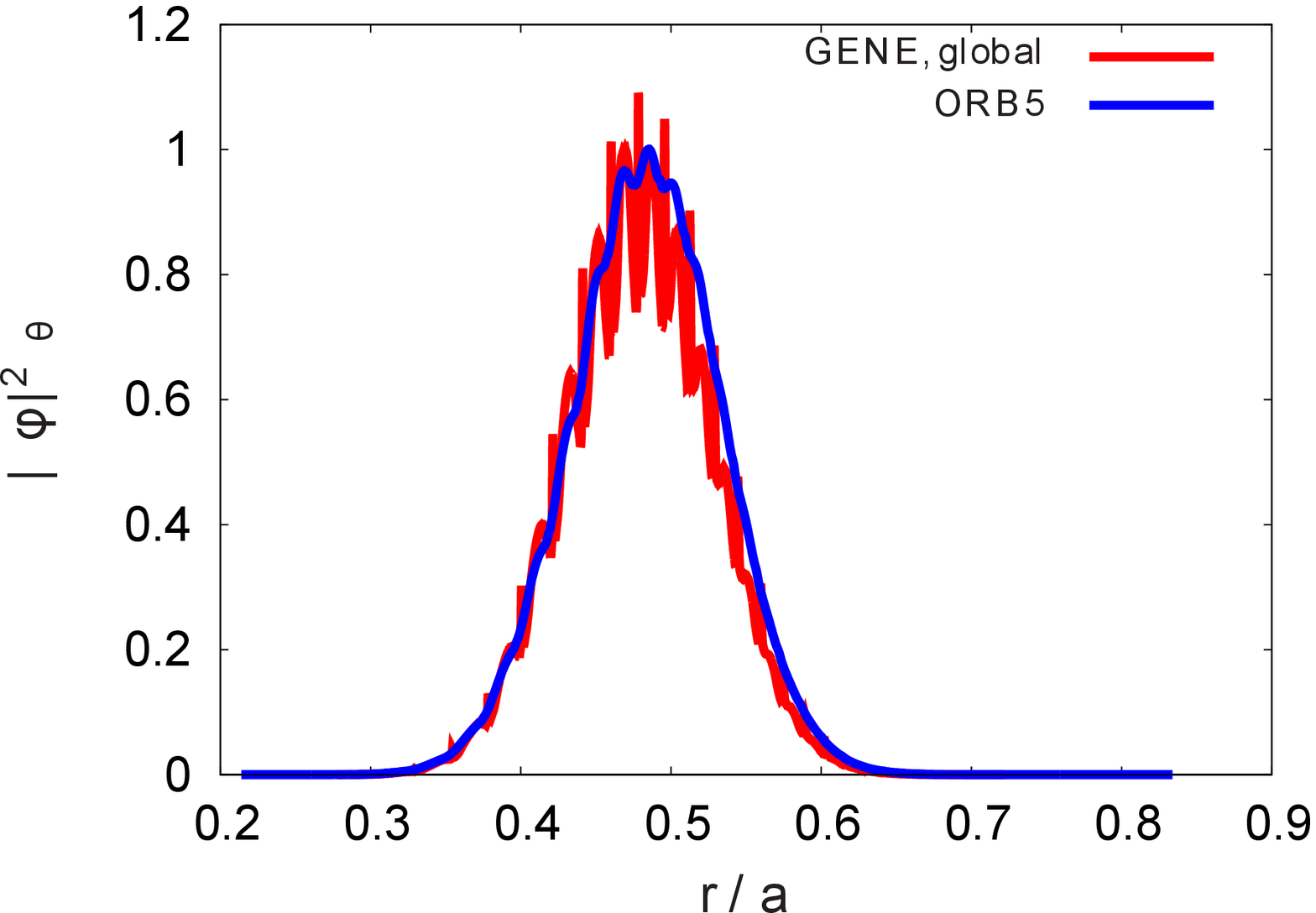}
\end{minipage}
 &
\begin{minipage}{.25\textwidth}
\includegraphics[width=\linewidth, height=3cm]{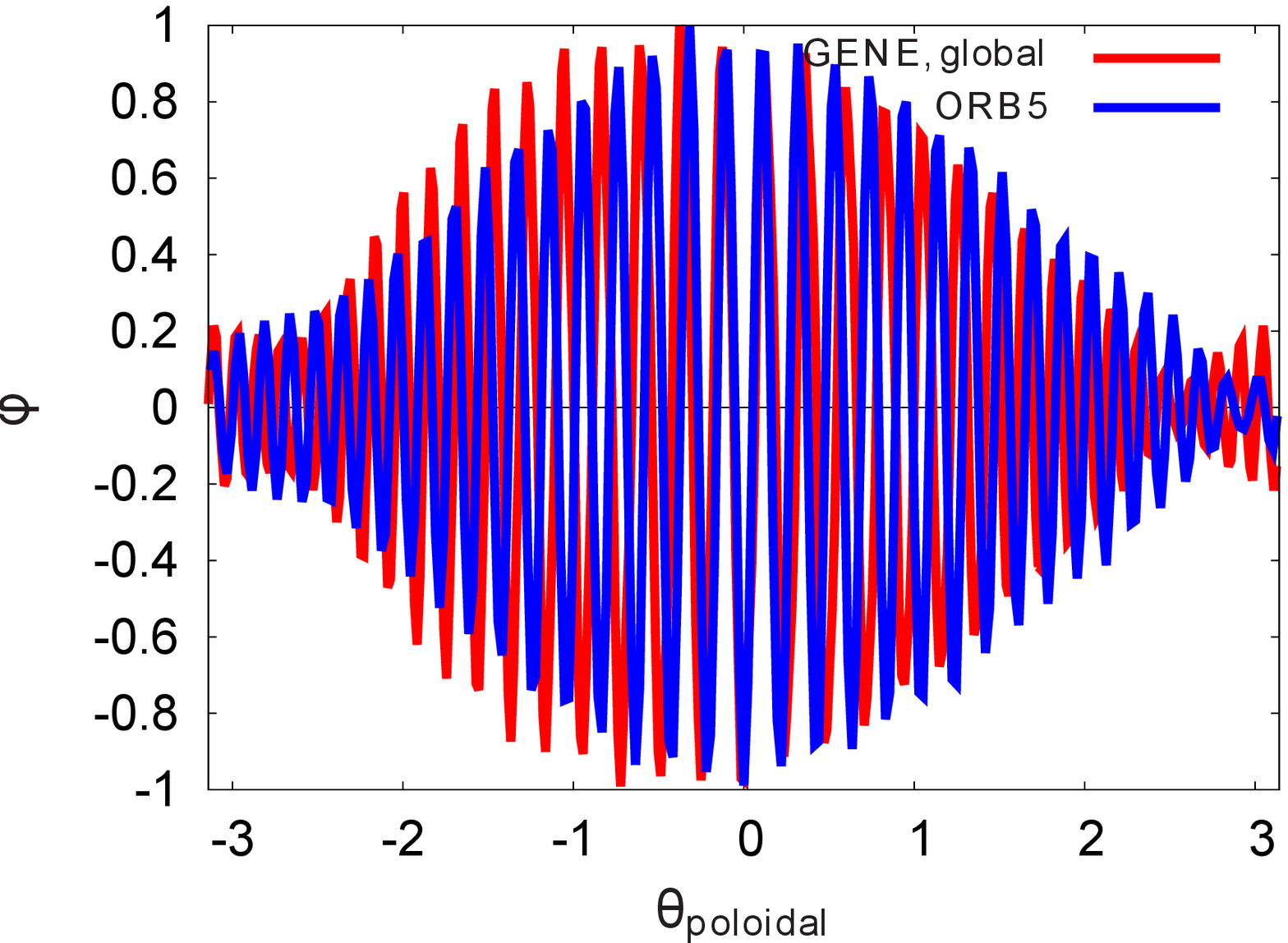}
\end{minipage}
\end{tabular}
\\
\begin{tabular}{c c}
\begin{minipage}{.25\textwidth}
\includegraphics[width=\linewidth, height=3cm]{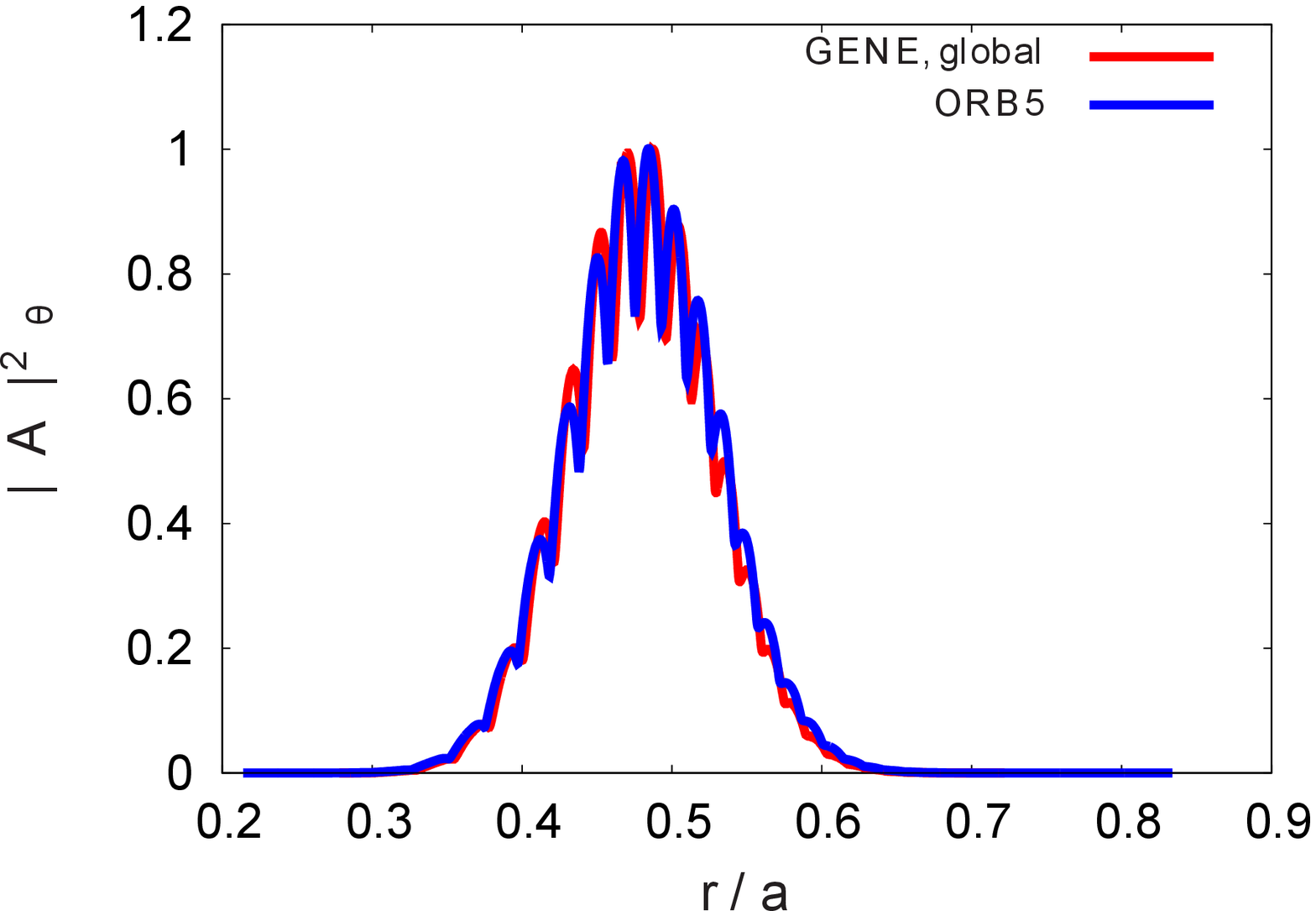}
\end{minipage}
 &
\begin{minipage}{.25\textwidth}
\includegraphics[width=\linewidth, height=3cm]{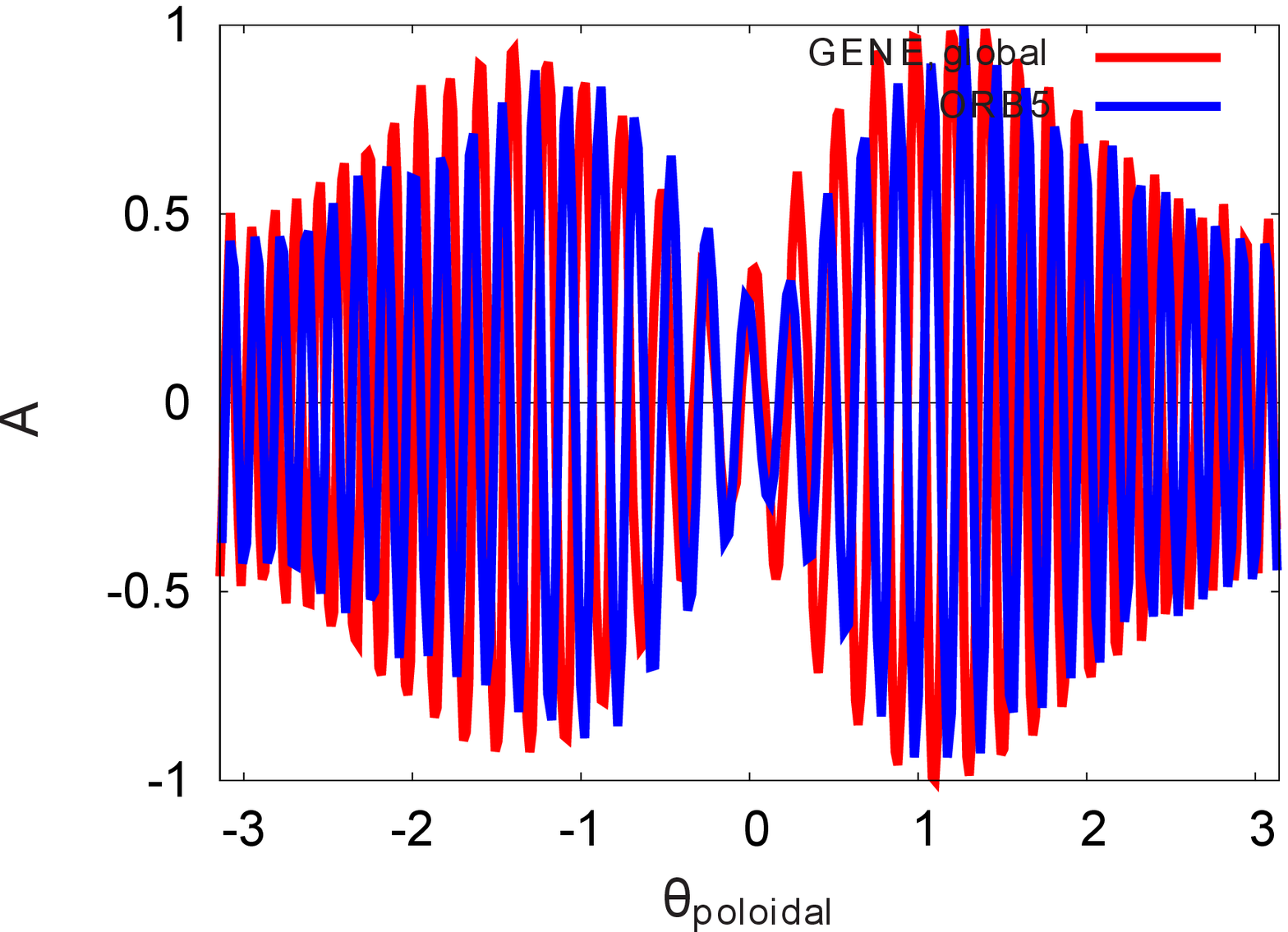}
\end{minipage}
\end{tabular}
\caption{Radial profiles of electrostatic and magnetic potentials: nominal $\beta=1\%$ and $n=25$.}
\label{radial_profiles}
\end{figure}
\section{\label{conclusions}Conclusions}
The two fold verification framework for gyrokinetic codes has been established. First of all, the models currently implemented in ORB5 (PIC representative code) and GENE (Eulerian representative code) have been derived within the Lagrangian variational framework, which permitted to perform a close comparison of both models and to identify the corresponding approximations. From the theoretical point of view (i.e. before performing all the approximations for further numerical implementations), it has been identified that the linear models for both codes are identical. In addition to that, the approximations performed on the theoretical models before the discretization have been also stated. 

On the other hand, the numerical part of the verification framework has shown an excellent agreement in the linear electromagnetic $\beta$ scan test case.  Further comparison of the nonlinear models at the theoretical level as well as extension of the benchmark to nonlinear simulations is the part of our future work.

\section{Acknowledgment}
Authors would like to acknowledge Cristel Chandre for the advice and help during preparation of this manuscript.
This work has been carried out within the framework of
the EUROfusion Consortium and has received funding from
the Euratom research and training Programme No. 2014-
2018 under Grant Agreement No. 633053. The views and
opinions expressed herein do not necessarily reflect those of
the European Commission.



\end{document}